\documentclass{elsart}
\usepackage{graphicx}
\renewcommand{\vec}[1]{\mbox{\boldmath ${#1}$}}
\begin{document}
\begin{frontmatter}
\title{Spin-wave instability for parallel pumping in ferromagnetic thin
 films under oblique field}
\author{Kazue Kudo\corauthref{cor}}, 
\corauth[cor]{Corresponding author. Tel.: +81 6 6605 2768; 
 fax: +81 6 6605 2768.}
\ead{kudo@a-phys.eng.osaka-cu.ac.jp}
\author{Katsuhiro Nakamura}
\address{Department of Applied Physics, Osaka City University,
              Osaka 558-8585, Japan }
\begin{abstract}
Spin-wave instability for parallel pumping is studied theoretically.
The spin-wave instability threshold is calculated in ferromagnetic
 thin films under oblique field which has an oblique angle to the film
 plane. 
The butterfly curve of the threshold usually has a cusp at a certain
 value of the static external field. While the static field value
of the cusp point
 varies as the oblique angle changes, the general properties of
 the butterfly curve show little noteworthy change. 
 For very thin films, however, multiple cusps of the butterfly
 curve can appear due to different standing spin-wave modes, which
 indicate a novel feature for thin films under oblique field. 
\end{abstract}
\end{frontmatter}

\section{Introduction}

The first experiment for the parametric excitation of spin wave by
parallel pumping was given by Schl\"oman {\it et al.}~\cite{Schlomann}. 
In a parallel pumping experiment, a microwave magnetic field is applied
parallel to the external static field. When the microwave field
amplitude $h$ exceeds a certain spin-wave instability amplitude 
$h_{\rm crit}$, the parametric spin-wave excitation occurs.
The excited spin waves have half the pumping frequency.
The $h_{\rm crit}$ curve plotted against the static field is called
``butterfly curve''.
The spin-wave instability for parallel pumping is often investigated 
for the purpose to
study relaxation phenomena in ferromagnetic materials, since the
instability threshold depends on the balance between the driving power
and the damping of spin waves.

Theoretical studies have not been very successful to explain 
the instability threshold for parallel pumping:    
a combination of 
more than two equations (theories) is needed to obtain the butterfly
curve. In other words, 
even a qualitative explanation for the butterfly curve cannot be
successful so long as one tries to use only the Landau-Lifshitz (LL)
equation governing the magnetization dynamics.  
By contrast the Suhl instability
for perpendicular pumping has been well explained
theoretically~\cite{Suhl}. 
The Suhl instability can be described by using only the LL equation. 
For parallel pumping, however, the LL equation 
is not enough to explain the instability, 
and the spin-wave line width $\Delta H_k$
is necessary to describe the relaxation of spin-waves as well as the
instability.  

A typical butterfly curve for parallel pumping has a cusp at a certain
external static field, which can be
explained if the spin-wave line width $\Delta H_k$ is given. For
example, Patton {\it et al.} proposed some trial $\Delta H_k$ functions
and successfully explained butterfly curves~\cite{Patton}.
Such a typical feature of butterfly curves usually does not depend on the
shape of samples. For very thin films, however, more interesting features
can appear. In fact, butterfly curves with multiple cusps were
observed for $0.5$-$\mu$m yttrium iron garnet (YIG) films under in-plane
external field~\cite{Kalin}. Those cusps were due to
quantized standing-wave modes across the film. 
It is highly desirable to develop the theory of multi-cusp butterfly
curves in the case of a general oblique field.

In this paper, we examine the instability threshold for parallel
pumping in ferromagnetic thin films  theoretically. 
The butterfly curve of the threshold will be calculated in cases where
the external field has an oblique angle to the film plane. 
For the calculation of the threshold, 
we use a trial $\Delta H_k$ function proposed by
Patton {\it et. al.}~\cite{Patton},
which was proposed originally for spherical samples but can be applied
to thin films~\cite{Wiese,Kabos}.
For usual thin films, the butterfly curve of the threshold has a single cusp at
a certain value of 
the static external field. We will show how the cusp shifts when the
oblique angle changes. Moreover, we will reveal that multi-cusp butterfly
curves can be seen for very thin films.

\section{Equations of motion for the spin-wave amplitudes}

The dynamics of magnetization field $\vec{M}(\vec{r})$ is governed by
the Landau-Lifshitz (LL) equation,
\begin{equation}
\frac1{\gamma}\partial_t \vec{M} =\vec{M}\times\vec{H}_{\rm eff}
 -\frac{\lambda}{M_0}\vec{M}\times (\vec{M}\times\vec{H}_{\rm eff}).
 \label{eq:LL}
\end{equation}
Here, $\gamma$ is the gyromagnetic ratio ($\gamma <0$ for spins);
 $M_0$ is the value of the
magnetization in thermal equilibrium; $\lambda$ is a damping constant
parameter ($\lambda <0$ in this case); 
$\vec{H}_{\rm eff}$ is the effective magnetic field:
\begin{equation}
\vec{H}_{\rm eff}=D\nabla^2\vec{M} +\vec{H}^{\rm d}+\vec{H}^{\rm a} 
+\vec{H}_0 +\vec{h}\cos\omega t.
\label{eq:Heff}
\end{equation}  
The first term comes from the exchange interaction; the third term is
the anisotropy field, which is omitted below for convenience;
the fourth and fifth
terms are the external static and pumping fields, respectively. For
parallel pumping, $\vec{H}_0 // \vec{h}$ and they are parallel to the
$z$ axis.
The second term on right hand side of Eq.~(\ref{eq:Heff}) is the
demagnetizing filed given by the gradient of the magneto-static potential
$\phi$: 
\begin{equation}
 \vec{H}^{\rm d}=-\vec{\nabla}\phi .
\end{equation}
The magneto-static potential obeys the Poisson equation:
\begin{equation}
 \nabla^2\phi = \left\{ 
\begin{array}{rl}
 4\pi\vec{\nabla}\cdot\vec{M}, & \mbox{inside the sample}, \\
 0, & \mbox{outside the sample}.
\end{array}
\right.
\label{eq:poi}
\end{equation}

The number of independent components of $\vec{M}$ is two, since the
length of the magnetization vector is invariant ($|\vec{M}|=M_0$).
The normalized magnetization $\vec{S}=\vec{M}/M_0$ can be
represented by a point on a unit sphere.
It is convenient to project the unit sphere stereographically onto a
complex variable $\psi (\vec{r},t)$~\cite{Laksh}:
\begin{equation}
 \psi=\frac{S_x+\mathrm{i}S_y}{1+S_z},
\end{equation}
where
\begin{equation}
 S_x=\frac{\psi+\psi^*}{1+\psi\psi^*}, \quad 
 S_y=\frac{\mathrm{i}(\psi^* -\psi)}{1+\psi\psi^*}, \quad
 S_z=\frac{1-\psi\psi^*}{1+\psi\psi^*}.
\label{eq:Ss}
\end{equation}
In terms of $\psi$, the LL equation~(\ref{eq:LL}) is rewritten as 
\begin{eqnarray}
 \partial_t \psi -\mathrm{i}(1-\mathrm{i}\lambda )&\gamma&
 \left\{ D M_0 \left[ \nabla^2\psi 
 -\frac{2\psi^*(\nabla\psi)^2}{1+\psi\psi^*} \right]
 -\frac12 (1-\psi^2)\partial_x\phi \right. \nonumber\\
 &-&\left. \frac{\rm i}2 (1+\psi^{*2})\partial_y\phi 
 +(\partial_z\phi-H_0-h\cos\omega t)\psi 
\right\} =0.
\label{eq:LLs}
\end{eqnarray}
Inside the sample, $\phi$ satisfies
\begin{eqnarray}
 \nabla^2\phi&=&\frac{4\pi M_0}{(1+\psi\psi^*)^2}\left\{
 (1-\psi^{*2})\partial_x\psi+(1-\psi^2)\partial_x\psi^* \right.\nonumber\\
 &+&\left. \mathrm{i}[(1+\psi^2)\partial_y\psi^*-(1+\psi^{*2})\partial_y\psi ]
 -2(\psi\partial_z\psi^* +\psi^*\partial_z\psi) \right\}.
\label{eq:pois}
\end{eqnarray}

\begin{figure}
\begin{center}
 \includegraphics[width=8cm]{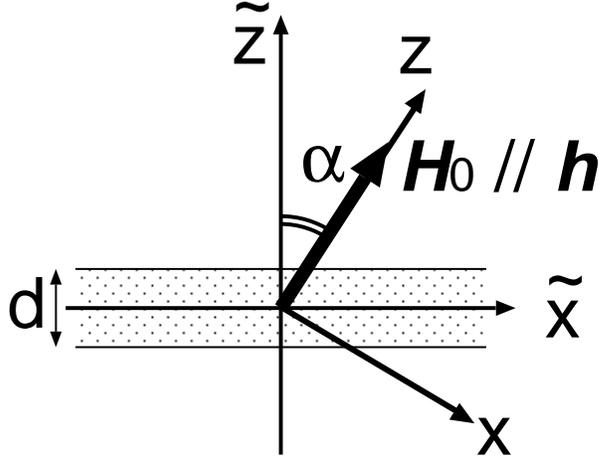}
\end{center}
\caption{Schematic picture of sample setting. The sample extends on the
 $\tilde{x}$-$\tilde{y}$ plane. The $z$ axis corresponds to the direction of
 the external filed and has an angle $\alpha$ to the $\tilde{z}$ axis.}
\label{fig:setting}
\end{figure}
Now let us consider the boundary conditions, which affect the
demagnetizing field. We assume a film of thickness $d$ infinitely
extended in the $\tilde{x}$-$\tilde{y}$ plane as shown in
Fig.~\ref{fig:setting}. The external field is at an angle $\alpha$ to
the $\tilde{z}$ axis and corresponds to the $z$ axis.
We assume unpinned surface spins, which satisfy Neumann-like boundary
conditions, 
\begin{equation}
 \partial_{\tilde{z}} \vec{S} |_{\tilde{z}=\pm d/2} =0,
\end{equation}
namely,
\begin{equation}
  \left. \frac{\partial}{\partial\tilde{z}}\psi
  \right|_{\tilde{z}=\pm d/2}
 =\left. \frac{\partial}{\partial\tilde{z}}\psi^*
  \right|_{\tilde{z}=\pm d/2}=0.
\label{eq:BC}
\end{equation}

Before proceeding to the linear instability of spin waves, we introduce the
following dimensionless time and space units~\cite{Elmer}:
\begin{equation}
 t\to \frac{t}{4\pi |\gamma |M_0}, \quad \vec{r}\to\vec{r}d.
\end{equation}
Then we obtain the linearized equations of motion corresponding to
Eqs.~(\ref{eq:LLs}) and (\ref{eq:pois}): 
\begin{equation}
\partial_t\psi +\mathrm{i}(1-\mathrm{i}\lambda)\left[
l^2\nabla^2\psi -\frac12(\partial_x +\mathrm{i}\partial_y)\Phi
+(\partial_z\Phi -\omega_H -\omega_h\cos\omega_{\rm p}t)\psi
\right] =0,
\label{eq:LLl} 
\end{equation}
\begin{equation}
 \nabla^2\Phi =(\partial_x -\mathrm{i}\partial_y)\psi
 +(\partial_x +\mathrm{i}\partial_y)\psi^* \quad 
 \mbox{when $-\frac12 < \tilde{z} < \frac12$},
\label{eq:poil} 
\end{equation}
where
\begin{equation}
 l^2=\frac{D}{4\pi d^2}, \quad \Phi =\frac{\phi}{4\pi M_0d}, \quad
 \omega_H =\frac{H_0}{4\pi M_0}, \quad \omega_h =\frac{h}{4\pi M_0},
 \quad \omega_{\rm p}=\frac{\omega}{4\pi M_0|\gamma |}.
\end{equation}

Now we consider the undriven case (i.e., $\omega_h=0$),
and expand $\psi(\vec{r},t)$ and $\psi^*(\vec{r},t)$ 
so that they fulfills the boundary
conditions~(\ref{eq:BC}). For even modes, $\tilde{k}_z=2m\pi$ ($m$:
integer),
\begin{eqnarray}
 \psi (\tilde{\vec{r}},t)&=&\sum_{\tilde{\vec{\scriptstyle k}}}
 a_{\tilde{\vec{\scriptstyle k}}}(t) 
 \e^{\mathrm{i}(\tilde{k}_x\tilde{x}+\tilde{k}_y\tilde{y})}
 \cos\tilde{k}_z\tilde{z} \nonumber\\
 \psi^* (\tilde{\vec{r}},t)&=&\sum_{\tilde{\vec{\scriptstyle k}}}
 a^*_{-\tilde{\vec{\scriptstyle k}}}(t) 
 \e^{\mathrm{i}(\tilde{k}_x\tilde{x}+\tilde{k}_y\tilde{y})}
 \cos\tilde{k}_z\tilde{z}. 
\label{eq:expan_a}
\end{eqnarray}
For odd modes, $\tilde{k}_z=(2m+1)\pi$ ($m$: integer),
\begin{eqnarray}
 \psi (\tilde{\vec{r}},t)&=&\sum_{\tilde{\vec{\scriptstyle k}}}
 a_{\tilde{\vec{\scriptstyle k}}}(t) 
 \e^{\mathrm{i}(\tilde{k}_x\tilde{x}+\tilde{k}_y\tilde{y})}
 \sin\tilde{k}_z\tilde{z} \nonumber\\
 \psi^* (\tilde{\vec{r}},t)&=&-\sum_{\tilde{\vec{\scriptstyle k}}}
 a^*_{-\tilde{\vec{\scriptstyle k}}}(t) 
 \e^{\mathrm{i}(\tilde{k}_x\tilde{x}+\tilde{k}_y\tilde{y})}
 \sin\tilde{k}_z\tilde{z}. 
\label{eq:expan_b}
\end{eqnarray}
Here, $\tilde{k}_x=k_x\cos\alpha +k_z\sin\alpha$, $\tilde{k}_y=k_y$, and
$\tilde{k}_z=k_z\cos\alpha -k_x\sin\alpha$.
Using the expansions~(\ref{eq:expan_a}) and (\ref{eq:expan_b}), we obtain 
solutions of Eq.~(\ref{eq:poil}): for even modes,
\begin{eqnarray}
 \Phi (\tilde{\vec{r}})&=&-\sum_{\tilde{\vec{\scriptstyle k}}}
 \frac{\e^{\mathrm{i}(\tilde{k}_x\tilde{x}+\tilde{k}_y\tilde{y})}}
 {\tilde{k}^2} \left[
\mathrm{i}(\tilde{k}_{-}a_{\tilde{\vec{\scriptstyle k}}}
+\tilde{k}_{+}a^*_{-\tilde{\vec{\scriptstyle k}}})
\left\{ \cos\tilde{k}_z\tilde{z}-(-1)^m\e^{-\tilde{k}_\perp /2}
\cosh\tilde{k}_\perp\tilde{z}  \right\} \right. \nonumber\\
&+& \left. \tilde{k}_z\sin\alpha\cdot (a_{\tilde{\vec{\scriptstyle k}}} +
a^*_{-\tilde{\vec{\scriptstyle k}}})\left\{
\sin\tilde{k}_z\tilde{z}-(-1)^m\e^{-\tilde{k}_\perp /2}
(\tilde{k}_z/\tilde{k}_\perp )
\sinh\tilde{k}_\perp\tilde{z} \right\}
\right],
\end{eqnarray}
for odd modes,
\begin{eqnarray}
  \Phi (\tilde{\vec{r}})&=&-\sum_{\tilde{\vec{\scriptstyle k}}}
 \frac{\e^{\mathrm{i}(\tilde{k}_x\tilde{x}+\tilde{k}_y\tilde{y})}}
 {\tilde{k}^2} \left[
\mathrm{i}(\tilde{k}_{-}a_{\tilde{\vec{\scriptstyle k}}}
-\tilde{k}_{+}a^*_{-\tilde{\vec{\scriptstyle k}}})
\left\{ \sin\tilde{k}_z\tilde{z}-(-1)^m\e^{-\tilde{k}_\perp /2}
\sinh\tilde{k}_\perp\tilde{z}  \right\} \right. \nonumber\\
&-& \left. \tilde{k}_z\sin\alpha\cdot (a_{\tilde{\vec{\scriptstyle k}}} -
a^*_{-\tilde{\vec{\scriptstyle k}}})\left\{
\cos\tilde{k}_z\tilde{z}+(-1)^m\e^{-\tilde{k}_\perp /2}
(\tilde{k}_z/\tilde{k}_\perp )
\cosh\tilde{k}_\perp\tilde{z} \right\}
\right],
\end{eqnarray}
where $\tilde{k}_{+}=\tilde{k}_x\cos\alpha +\mathrm{i}\tilde{k}_y$, 
$\tilde{k}_{-}=\tilde{k}_x\cos\alpha -\mathrm{i}\tilde{k}_y$ and
$\tilde{k}_\perp =\sqrt{\tilde{k}_x^2+\tilde{k}_y^2}$.
The detailed derivation of these solutions is shown in
Appendix~\ref{sec:app}.
Here we approximate $\partial_z\Phi=-H^{\rm d}_z/4\pi M_0$ 
with the value of uniform magnetization, $\vec{k}=0$: 
$H^{\rm d}_z=H^{\rm d}_{\tilde{x}}\sin\alpha 
+H^{\rm d}_{\tilde{z}}\cos\alpha =-4\pi 
(N_{\tilde{x}}M_{\tilde{x}}\sin\alpha 
+N_{\tilde{z}} M_{\tilde{z}}\cos\alpha)$, where $N_{\tilde{x}}$ and 
$N_{\tilde{z}}$ are
demagnetizing factors. In this case, $N_{\tilde{x}}=N_{\tilde{y}}=0$ and
$N_{\tilde{z}}=1$ .
Assuming $M_z\simeq M_0$, namely, 
$M_{\tilde{z}}\simeq M_0\cos\alpha$, we obtain  
$\partial_z\Phi =-H^{\rm d}_z/4\pi M_0=\cos^2 \alpha$.
Then Eq.~(\ref{eq:LLl}) for $\omega_h=0$ is rewritten as
\begin{equation}
\partial_t\psi +\mathrm{i}(1-\mathrm{i}\lambda)\left[
l^2\nabla^2\psi -\frac12(\cos\alpha\partial_{\tilde{x}} 
 +\mathrm{i}\partial_{\tilde{y}}-\sin\alpha\partial_{\tilde{z}})\Phi
+(\cos^2 \alpha -\omega_H )\psi \right] =0.
\label{eq:LLl2}
\end{equation}
Combining Eqs.~(\ref{eq:expan_a})-(\ref{eq:LLl2}), we obtain
\begin{equation}
\partial_t a_{\vec{\scriptstyle k}}
 +\mathrm{i}(1-\mathrm{i}\lambda) A_{\vec{\scriptstyle k}}
   a_{\vec{\scriptstyle k}}
 +\mathrm{i}(1-\mathrm{i}\lambda) B_{\vec{\scriptstyle k}}
   a^*_{-\vec{\scriptstyle k}}=0, 
\label{eq:al}
\end{equation}
where
\begin{equation}
 A_{\vec{\scriptstyle k}}= C_{\vec{\scriptstyle k}}+ 
 \mathrm{Re}\tilde{C_{\vec{\scriptstyle k}}}, \quad
 B_{\vec{\scriptstyle k}}= (-1)^n\left(
D_{\vec{\scriptstyle k}}
\e^{2\mathrm{i}\varphi_{\vec{\scriptscriptstyle k}}}
+\tilde{C_{\vec{\scriptstyle k}}}
\right). 
\end{equation}
Here, $n$ is an integer, 
$\cos\varphi_{\vec{\scriptstyle k}}=k_x/k_\perp$,
$\sin\varphi_{\vec{\scriptstyle k}}=k_y/k_\perp$,
$k_{\perp}=\sqrt{k_x^2+k_y^2}$, and
\begin{eqnarray}
C_{\vec{\scriptstyle k}}&=&\cos^2 \alpha -l^2 k^2 -\omega_H-\frac12 
f_{\vec{\scriptstyle k}}\sin^2\theta_{\vec{\scriptstyle k}},\nonumber\\
\tilde{C_{\vec{\scriptstyle k}}}&=&-\frac{\tilde{k}_z}{k^2}\sin\alpha
\left( k_{\perp}\e^{\mathrm{i}\varphi_{\vec{\scriptscriptstyle k}}}
+\tilde{k}_z\sin\alpha \right) f_{\vec{\scriptstyle k}},\nonumber\\
D_{\vec{\scriptstyle k}}&=&-\frac12 f_{\vec{\scriptstyle k}}
\sin^2\theta_{\vec{\scriptstyle k}},\nonumber\\
f_{\vec{\scriptstyle k}}&=&1-(2-\delta_{0n})\left(
1-\exp \left[ -\sqrt{k^2-\tilde{k}_z^2}\right] \right)
\frac{\sqrt{k^2-\tilde{k}_z^2}}{k^2},\nonumber\\
&&\sin\theta_{\vec{\scriptstyle k}}=\frac{k_\perp}{k}, \quad 
\tilde{k}_z =\pi n.
\label{eq:CDf}
\end{eqnarray} 
The detailed derivation is given in Appendix~\ref{sec:app_b}. 

Equation~(\ref{eq:al}) represents two coupled harmonic oscillators 
$a_{\vec{\scriptstyle k}}$, $a^*_{-\vec{\scriptstyle k}}$.
We diagonalize Eq.~(\ref{eq:al}) by means of the Holstein-Primakoff
transformation: 
\begin{eqnarray}
 a_{\vec{\scriptstyle k}}&=&
 \nu_{\vec{\scriptstyle k}}b_{\vec{\scriptstyle k}}
-\mu_{\vec{\scriptstyle k}}b^*_{-\vec{\scriptstyle k}}\nonumber\\
 a^*_{-\vec{\scriptstyle k}}&=&
 \nu_{\vec{\scriptstyle k}}b^*_{-\vec{\scriptstyle k}}
-\mu^*_{\vec{\scriptstyle k}}b_{\vec{\scriptstyle k}}, 
\label{eq:H-P}
\end{eqnarray}
where
\begin{eqnarray}
  \nu_{\vec{\scriptstyle k}}&=&
 \cosh\frac{\chi_{\vec{\scriptstyle k}}}{2},\nonumber\\
 \mu_{\vec{\scriptstyle k}}&=&
 \e^{\mathrm{i}\beta_{\vec{\scriptscriptstyle k}}}
 \sinh\frac{\chi_{\vec{\scriptstyle k}}}{2},\nonumber\\
 \cosh\chi_{\vec{\scriptstyle k}}&=&
 \frac{\left| C_{\vec{\scriptstyle k}}+
 \mathrm{Re}\tilde{C_{\vec{\scriptstyle k}}}\right| }
{\left[ (C_{\vec{\scriptstyle k}}+
 \mathrm{Re}\tilde{C_{\vec{\scriptstyle k}}})^2-(1+\lambda^2)
\left| D_{\vec{\scriptstyle k}}
 \e^{2\mathrm{i}\varphi_{\vec{\scriptscriptstyle k}}}+
\tilde{C_{\vec{\scriptstyle k}}}\right|^2 \right]^{1/2}}.
\end{eqnarray}
Substituting Eq.~(\ref{eq:H-P}) into Eq.~(\ref{eq:al}), we obtain
\begin{equation}
 \partial_t b_{\vec{\scriptstyle k}} -\mathrm{i}
(\omega_{\vec{\scriptstyle k}}+\mathrm{i}\eta_{\vec{\scriptstyle k}})
 b_{\vec{\scriptstyle k}}=0, 
\end{equation}
where
\begin{eqnarray}
 \omega_{\vec{\scriptstyle k}}^2 &=&(C_{\vec{\scriptstyle k}}+
 \mathrm{Re}\tilde{C_{\vec{\scriptstyle k}}})^2
 -(1+\lambda^2) \left| D_{\vec{\scriptstyle k}}
 \e^{2\mathrm{i}\varphi_{\vec{\scriptscriptstyle k}}}+
 \tilde{C_{\vec{\scriptstyle k}}}\right|^2,
 \label{eq:disp} \\
 \eta_{\vec{\scriptstyle k}}&=&\lambda (C_{\vec{\scriptstyle k}}+
 \mathrm{Re}\tilde{C_{\vec{\scriptstyle k}}})
\label{eq:damp}
\end{eqnarray}
Equations~(\ref{eq:disp}) and (\ref{eq:damp}) express the dispersion
relation and a damping rate, respectively.

\section{The instability threshold}

Now we consider the case where the microwave field $\vec{h}\cos\omega t$
is applied. The equation of motion of $a_{\vec{\scriptstyle k}}$
corresponding to Eq.~(\ref{eq:al}) is 
\begin{equation}
 \partial_t a_{\vec{\scriptstyle k}}
 +\mathrm{i}(1-\mathrm{i}\lambda) A_{\vec{\scriptstyle k}}
   a_{\vec{\scriptstyle k}}
 +\mathrm{i}(1-\mathrm{i}\lambda) B_{\vec{\scriptstyle k}}
   a^*_{-\vec{\scriptstyle k}}
 -\mathrm{i}(1-\mathrm{i}\lambda) \omega_h\cos\omega_{\rm p}t\cdot
  a_{\vec{\scriptstyle k}}=0. 
\label{eq:ap}
\end{equation}
After the the Holstein-Primakoff transformation (\ref{eq:H-P}), 
we substitute the following equations into Eq.~(\ref{eq:ap}),
\begin{eqnarray}
 b_{\vec{\scriptstyle k}}(t)&=&b^o_{\vec{\scriptstyle k}}(t)
\exp [\mathrm{i}(\omega_{\rm p}/2)t-\eta_{\vec{\scriptstyle k}}t],
\nonumber\\
 b^*_{-\vec{\scriptstyle k}}(t)&=&b^{o*}_{-\vec{\scriptstyle k}}(t)
\exp [-\mathrm{i}(\omega_{\rm p}/2)t-\eta_{\vec{\scriptstyle k}}t],
\end{eqnarray}
since the resonance occurs at 
$\omega_{\vec{\scriptstyle k}}=\omega_{\rm p}/2$.
Then the slowly-varying variable $b^o_{\vec{\scriptstyle k}}$ satisfies
\begin{equation}
 \partial_t^2 b^o_{\vec{\scriptstyle k}} +
 \left[ \left( \omega_{\vec{\scriptstyle k}} 
 -\frac{\omega_{\rm p}}{2}\right)^2 
 - \left| \rho_{\vec{\scriptstyle k}} \right|^2 \right]
 b^o_{\vec{\scriptstyle k}}=0,
\label{eq:bp}
\end{equation} 
where
\begin{equation}
 \left|  \rho_{\vec{\scriptstyle k}} \right| =
 \omega_h\sqrt{1+\lambda^2}\frac{\left| D_{\vec{\scriptstyle k}}
 \e^{2\mathrm{i}\varphi_{\vec{\scriptscriptstyle k}}}+
 \tilde{C_{\vec{\scriptstyle k}}}\right| }
 {2\omega_{\vec{\scriptstyle k}}}. 
\label{eq:rho}
\end{equation}
Therefore, the exponentially increasing solution for 
$b_{\vec{\scriptstyle k}}$ is 
\begin{equation}
 b_{\vec{\scriptstyle k}}\propto \exp \left[
\left( \left| \rho_{\vec{\scriptstyle k}} \right| 
-\eta_{\vec{\scriptstyle k}} \right)t +\mathrm{i}(\omega_{\rm p}/2)
t \right],
\end{equation}
where
\begin{equation}
 \left| \rho_{\vec{\scriptstyle k}} \right| > \eta_{\vec{\scriptstyle k}}.
\end{equation}
The instability threshold $\omega_h^{\rm crit}$ is now given as
\begin{equation}
 \omega_h^{\rm crit}=\frac{\omega_{\rm p}}{\sqrt{1+\lambda^2}} 
 \min_{\vec{\scriptstyle k}} \left\{ 
 \frac{\eta_{\vec{\scriptstyle k}}}{\left| D_{\vec{\scriptstyle k}}
 \e^{2\mathrm{i}\varphi_{\vec{\scriptscriptstyle k}}}+
 \tilde{C_{\vec{\scriptstyle k}}}\right| } \right\}.
\label{eq:dam_dH}
\end{equation}
The threshold, Eq.~(\ref{eq:dam_dH}), is obtained by using only the LL
equation, but cannot explain the experimentally-observed instability
threshold for parallel pumping.
In fact, Eq.~(\ref{eq:dam_dH}) proves to give an instability curve
totally different from the real butterfly curve.
Since the relaxation phenomenon is essentially nonlinear one, 
the linearizion analysis is not sufficient to discuss the instability
threshold. The problem is beyond the purpose of this paper, and 
it will be discussed in a forthcoming paper~\cite{Kudo}.
This difficulty, however, can be overcome by replacing the damping rate 
$\eta_{\vec{\scriptstyle k}}$ by a suitable spin-wave line width.
Using the spin-wave line width $\Delta H_{\vec{\scriptstyle k}}$, 
we rewrite Eq.~(\ref{eq:dam_dH}) as
\begin{equation}
  \omega_h^{\rm crit}=\omega_{\rm p}
 \min_{\vec{\scriptstyle k}} \left\{ 
 \frac{\Delta H_{\vec{\scriptstyle k}}}{\left| D_{\vec{\scriptstyle k}}
 \e^{2\mathrm{i}\varphi_{\vec{\scriptscriptstyle k}}}
 \tilde{C_{\vec{\scriptstyle k}}}\right| } \right\}.
\label{eq:dH}
\end{equation}
Here we adopt a simple trial $\Delta H_{\vec{\scriptstyle k}}$
function~\cite{Patton},
\begin{equation}
 \Delta  H_{\vec{\scriptstyle k}}=A_0+A_1\sin^2(2\theta_k) +A_2k,
\label{eq:dH2}
\end{equation}
where $A_0$, $A_1$ and $A_2$ are adjustable parameters.

\section{Butterfly curves}

Typical butterfly curves of the threshold $\omega_h^{\rm crit}$ 
have a cusp at a
certain static field: as the static field increasing, $\omega_h^{\rm crit}$ 
decreases
below the cusp point and increases above that.
For static fields below the cusp point, the minimum threshold modes corresponds
to a spin wave propagating with $\theta_{\vec{\scriptstyle k}}=\pi /2$.
As the static field increases, the wave vector $k$ of the threshold
modes decreases, and $k \simeq 0$ at the cusp.
For static fields above the cusp point, the wave number remains at $k \simeq 0$
and $\theta_{\vec{\scriptstyle k}}$ decreases from $\pi /2$ to $0$.

\begin{figure}
 \begin{center}
  \includegraphics[width=14cm]{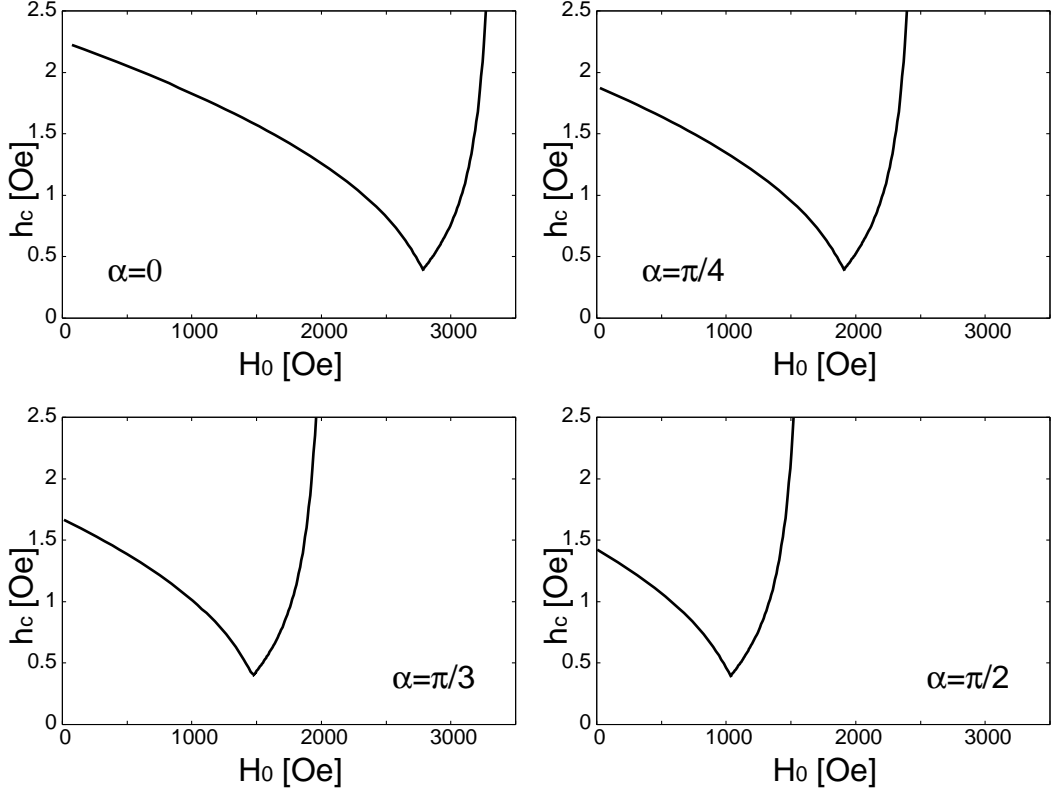}
 \end{center}
\caption{Theoretical butterfly curves for $d=15\mu$m for some values of
 the oblique angle $\alpha$.}
\label{fig:15m}
\end{figure}
Figure~\ref{fig:15m} shows theoretical butterfly curves calculated by
using 
Eqs.~(\ref{eq:dH}) and (\ref{eq:dH2}).  
The material parameters used in the calculation 
are typical values for
yttrium iron garnet (YIG) materials: 
$|\gamma |=1.77\times 10^7$rad/(s$\cdot$Oe), $4\pi M_0=1.75\times10^3$Oe,
$D=5.4\times 10^{-9}$Oe$\cdot$cm$^2$/rad$^2$. 
The other parameters are as follows: 
$\omega /(2\pi )=9.5\times 10^9$Hz, $d=15\mu$m, 
$A_0=3.0\times 10^{-2}/(4\pi M_0)$,
$A_1/A_0=0.4$, $A_2/A_0=3.0\times 10^{-4}/d$.  
We calculate the static field value for the cusp point with $k=0$ and 
$\theta_{\vec{\scriptstyle k}}$ in a bulk
approximation: we assume $f_{\vec{\scriptstyle k}}=1$ in
Eq.~(\ref{eq:CDf}). Setting 
$\omega_{\vec{\scriptstyle k}}=\omega_{\rm p}/2$ in Eq.~(\ref{eq:disp}),
we obtain the static field for the cusp point
\begin{equation}
 \omega_H^{\rm crit}=\cos^2\alpha 
-\frac12+\frac12\sqrt{\omega_{\rm p}^2+1}.
\end{equation}
While the cusp point shifts to lower static fields as the oblique angle
$\alpha$ increases, the typical shape of curves is found to show little
noteworthy change.

In some cases, however, 
curious butterfly curves can emerge. 
\begin{figure}
 \begin{center}
  \includegraphics[width=7cm]{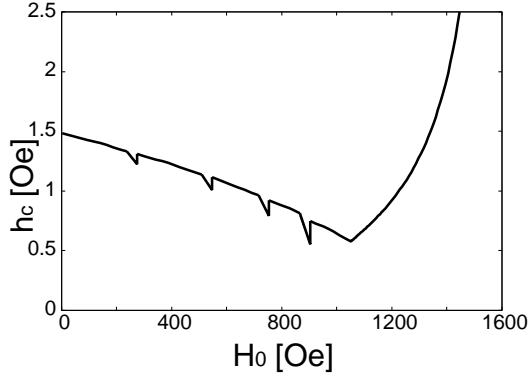}
 \end{center}
\caption{Theoretical butterfly curve for $d=5\mu$m. The external field
 is applied parallel to the film plane.}
\label{fig:5m}
\end{figure}
Figure~\ref{fig:5m} is a theoretical butterfly curve for the case of
$d=5\mu$m and $\alpha=\pi/2$. 
The other parameters used in the calculation are the same as used in 
Fig.~\ref{fig:15m}.
There appear multiple cusps in the low field region of the butterfly
curve. 
These cusps are due to different standing spin-wave modes across the
film thickness. 
Let us recall that the $\tilde{z}$ component of wave vectors are
quantized. Effects of quantization are essential when the boundary
conditions become important. Comparing the results in
Figs.~\ref{fig:15m} and \ref{fig:5m}, one may say 
multi-cusp feature of a butterfly curve can
be seen when the thickness of a film is very small. 
This finding is supported by some experimental
studies~\cite{Kalin,Wiese}. In Ref.~\cite{Wiese}, a standard butterfly
curve of the instability was shown for 15.9-$\mu$m-thick yttrium iron
garnet (YIG) film under in-plane external field. However, in
Ref.~\cite{Kalin}, multiple butterfly curves appeared for a YIG film
under in-plane external field. The difference between the two is just
the thickness. The thickness of the film in Ref.~\cite{Kalin} was
0.5$\mu$m.

\section{Discussion}

We have revealed how butterfly curves depend on the oblique angle between
the external field and the film plane. 
We have calculated theoretical butterfly curves by using the LL equation
together with the spin-wave line width  $\Delta H_k$.
The calculation was performed
under the assumption that parameters $A_0$, $A_1$ and $A_2$ for 
$\Delta H_k$ do not depend
on the oblique angle.  
These parameters were originally introduced
to fit a theoretical butterfly curve to experimental
data~\cite{Patton}, and
might depend on the oblique angle.

We have also shown qualitative features of the novel aspect of butterfly
curves with multiple cusps for parallel pumping.
Those multi-cusp curves come from quantized standing-wave modes
across the film. However, it is not clear how such cusps appear, 
since there remains an ambiguity about how to
evaluate $A_0$, $A_1$ and $A_2$. 
Figure~\ref{fig:5m} is one of possible theoretical curves. 
To proceed to quantitative evaluations, the
experimental test is highly desirable to confirm the
features predicted here.

\section*{Acknowledgments}

The authors thank to Prof. Mino of Okayama university for useful
discussion. One of the authors (K. K.) is supported by JSPS Research
Fellowships for Young Scientists.

\appendix

\section{\label{sec:app} Demagnetization field}

First, we should use the $(\tilde{x},\tilde{y},\tilde{z})$ coordinate
system rather than $(x,y,z)$ 
to use Eqs.~(\ref{eq:expan_a}) and (\ref{eq:expan_b}) in the
Poisson equation~(\ref{eq:poi}), and employ the transformation:
$\nabla^2\to \tilde{\nabla}^2$, 
$\partial_x\to\cos\alpha\partial_{\tilde{x}}-\sin\alpha\partial_{\tilde{z}}$.
We consider only the case of $-\frac12 < \tilde{z} < \frac12$.
For even modes, $\tilde{k}_z=2m\pi$ ($m$: integer), the Poisson
equation is rewritten as 
\begin{eqnarray}
 \tilde{\nabla}^2\Phi =\sum_{\tilde{\vec{\scriptstyle k}}}
 \e^{\mathrm{i}(\tilde{k}_x\tilde{x}+\tilde{k}_y\tilde{y})}&&
 \left[ (\mathrm{i}\tilde{k}_{-}\cos\tilde{k}_z\tilde{z}
 +\tilde{k}_z\sin\alpha\sin\tilde{k}_z\tilde{z})
 a_{\vec{\scriptstyle k}}\right. \nonumber\\
 &+&\left. (\mathrm{i}\tilde{k}_{+}\cos\tilde{k}_z\tilde{z}
 +\tilde{k}_z\sin\alpha\sin\tilde{k}_z\tilde{z})
 a^*_{-\vec{\scriptstyle k}} \right],
\label{eq:A-poia}
\end{eqnarray}
where $\tilde{k}_{+}=\tilde{k}_x\cos\alpha +\mathrm{i}\tilde{k}_y$ and 
$\tilde{k}_{-}=\tilde{k}_x\cos\alpha -\mathrm{i}\tilde{k}_y$.
For odd modes, $\tilde{k}_z=(2m+1)\pi$ ($m$: integer),
\begin{eqnarray}
 \tilde{\nabla}^2\Phi =\sum_{\tilde{\vec{\scriptstyle k}}}
 \e^{\mathrm{i}(\tilde{k}_x\tilde{x}+\tilde{k}_y\tilde{y})}&&
 \left[ (\mathrm{i}\tilde{k}_{-}\sin\tilde{k}_z\tilde{z}
 -\tilde{k}_z\sin\alpha\cos\tilde{k}_z\tilde{z})
 a_{\vec{\scriptstyle k}}\right. \nonumber\\
 &-&\left. (\mathrm{i}\tilde{k}_{+}\sin\tilde{k}_z\tilde{z}
 -\tilde{k}_z\sin\alpha\cos\tilde{k}_z\tilde{z})
 a^*_{-\vec{\scriptstyle k}} \right].
\label{eq:A-poib}
\end{eqnarray}

Here we note a well known fact: the Poisson equation,
\begin{eqnarray}
 \nabla^2\phi(\vec{r})=-4\pi\rho (\vec{r}),
\end{eqnarray} 
has the solution as
\begin{eqnarray}
 \phi(\vec{r})=\int\frac{\rho (\vec{r}')\d
  \vec{V}'}{|\vec{r}'-\vec{r}|}.
\label{eq:a_phi}
\end{eqnarray}
Applying Eq.~(\ref{eq:a_phi}), 
we obtain the solution of Eqs.~(\ref{eq:A-poia}) 
and (\ref{eq:A-poib}): for even modes,
\begin{eqnarray}
 \Phi(\tilde{\vec{r}})=-\frac1{4\pi}&\sum_{\tilde{\vec{\scriptstyle k}}}&
\left\{
\mathrm{i}(\tilde{k}_{-}a_{\tilde{\vec{\scriptstyle k}}}+
 \tilde{k}_{+}a^*_{-\tilde{\vec{\scriptstyle k}}})
\int\frac{\d \tilde{\vec{V}}'}{|\tilde{\vec{r}}'-\tilde{\vec{r}}|}
\e^{\mathrm{i}(\tilde{k}_x\tilde{x}'+\tilde{k}_y\tilde{y}')}
\cos\tilde{k}_z\tilde{z}' \right. \nonumber\\
&+&\left. \tilde{k}_z\sin\alpha\cdot (a_{\tilde{\vec{\scriptstyle k}}}+
a^*_{-\tilde{\vec{\scriptstyle k}}} )
\int\frac{\d \tilde{\vec{V}}'}{|\tilde{\vec{r}}'-\tilde{\vec{r}}|}
\e^{\mathrm{i}(\tilde{k}_x\tilde{x}'+\tilde{k}_y\tilde{y}')}
\sin\tilde{k}_z\tilde{z}'
\right\};
\label{eq:a5}
\end{eqnarray}
for odd modes,
\begin{eqnarray}
 \Phi(\tilde{\vec{r}})=-\frac1{4\pi}&\sum_{\tilde{\vec{\scriptstyle k}}}&
\left\{
\mathrm{i}(\tilde{k}_{-}a_{\tilde{\vec{\scriptstyle k}}}-
 \tilde{k}_{+}a^*_{-\tilde{\vec{\scriptstyle k}}})
\int\frac{\d \tilde{\vec{V}}'}{|\tilde{\vec{r}}'-\tilde{\vec{r}}|}
\e^{\mathrm{i}(\tilde{k}_x\tilde{x}'+\tilde{k}_y\tilde{y}')}
\sin\tilde{k}_z\tilde{z}'\right.  \nonumber\\
&-&\left. \tilde{k}_z\sin\alpha\cdot (a_{\tilde{\vec{\scriptstyle k}}}-
a^*_{-\tilde{\vec{\scriptstyle k}}} )
\int\frac{\d \tilde{\vec{V}}'}{|\tilde{\vec{r}}'-\tilde{\vec{r}}|}
\e^{\mathrm{i}(\tilde{k}_x\tilde{x}'+\tilde{k}_y\tilde{y}')}
\cos\tilde{k}_z\tilde{z}'
\right\}.
\label{eq:a6}
\end{eqnarray}
After integration, Eqs.~(\ref{eq:a5}) and (\ref{eq:a6}) become
\begin{eqnarray}
 \Phi(\tilde{\vec{r}})=&-&\sum_{\tilde{\vec{\scriptstyle k}}}\frac
{\e^{\mathrm{i}(\tilde{k}_x\tilde{x}+\tilde{k}_y\tilde{y})}}
{\tilde{k}^2}
\left\{
\mathrm{i}(\tilde{k}_{-}a_{\tilde{\vec{\scriptstyle k}}}+
 \tilde{k}_{+}a^*_{-\tilde{\vec{\scriptstyle k}}})
\left[ \cos\tilde{k}_z\tilde{z}-(-1)^m\e^{-\tilde{k}_\perp /2}
\cosh\tilde{k}_\perp\tilde{z} \right]\right. \nonumber\\
&+&\left. \tilde{k}_z\sin\alpha\cdot (a_{\tilde{\vec{\scriptstyle k}}}+
a^*_{-\tilde{\vec{\scriptstyle k}}} )
\left[ \sin\tilde{k}_z\tilde{z}-(-1)^m\e^{-\tilde{k}_\perp /2}
\frac{\tilde{k}_z}{\tilde{k}_\perp}
\sinh\tilde{k}_\perp\tilde{z} \right]
\right\},
\end{eqnarray}
and
\begin{eqnarray}
 \Phi(\tilde{\vec{r}})=&-&\sum_{\tilde{\vec{\scriptstyle k}}}\frac
{\e^{\mathrm{i}(\tilde{k}_x\tilde{x}+\tilde{k}_y\tilde{y})}}
{\tilde{k}^2}
\left\{
\mathrm{i}(\tilde{k}_{-}a_{\tilde{\vec{\scriptstyle k}}}-
 \tilde{k}_{+}a^*_{-\tilde{\vec{\scriptstyle k}}})
\left[ \sin\tilde{k}_z\tilde{z}-(-1)^m\e^{-\tilde{k}_\perp /2}
\sinh\tilde{k}_\perp\tilde{z} \right]\right. \nonumber\\
&-&\left. \tilde{k}_z\sin\alpha\cdot (a_{\tilde{\vec{\scriptstyle k}}}-
a^*_{-\tilde{\vec{\scriptstyle k}}} )
\left[ \cos\tilde{k}_z\tilde{z}+(-1)^m\e^{-\tilde{k}_\perp /2}
\frac{\tilde{k}_z}{\tilde{k}_\perp}
\cosh\tilde{k}_\perp\tilde{z} \right]
\right\},
\end{eqnarray}
respectively.

\section{\label{sec:app_b} Equation of motion for $a_{\vec{\scriptstyle k}}$}

First, let us calculate the derivative of $\Phi$ in Eq.~(\ref{eq:LLl2}).
For even modes, $\tilde{k}_z=2m\pi$ ($m$: integer),
\begin{eqnarray}
 \lefteqn{(\cos\alpha\partial_{\tilde{x}}+\mathrm{i}\partial_{\tilde{y}}
-\sin\alpha\partial_{\tilde{z}})\Phi
=\sum_{\tilde{\vec{\scriptstyle k}}}
\frac{\e^{\mathrm{i}(\tilde{k}_x\tilde{x}+\tilde{k}_y\tilde{y})}}
{\tilde{k}^2}}
\nonumber\\
\times\biggl[  &-& \left\{
-(\tilde{k}_{+}\tilde{k}_{-}a_{\tilde{\vec{\scriptstyle k}}}
+\tilde{k}_{+}^2a^*_{-\tilde{\vec{\scriptstyle k}}})
\left[ \cos\tilde{k}_z\tilde{z}-(-1)^m\e^{-\tilde{k}_\perp /2}
\cosh\tilde{k}_\perp\tilde{z} \right] \right. \nonumber\\
&&+\left. \mathrm{i}\tilde{k}_{+}\tilde{k}_z\sin\alpha\cdot
(a_{\tilde{\vec{\scriptstyle k}}}+a^*_{-\tilde{\vec{\scriptstyle k}}})
\left[ \sin\tilde{k}_z\tilde{z}-(-1)^m\e^{-\tilde{k}_\perp /2}
(\tilde{k}_z /\tilde{k}_\perp )\sinh\tilde{k}_\perp\tilde{z} \right]
\right\} \nonumber\\
&+& \sin\alpha\left\{
\mathrm{i}(\tilde{k}_{-}a_{\tilde{\vec{\scriptstyle k}}}
+\tilde{k}_{+}a^*_{-\tilde{\vec{\scriptstyle k}}})
\left[ -\tilde{k}_z\sin\tilde{k}_z\tilde{z}-(-1)^m
\e^{-\tilde{k}_\perp /2}
\tilde{k}_\perp\sinh\tilde{k}_\perp\tilde{z} \right] 
\right. \nonumber\\
&&+\left. \tilde{k}_z^2\sin\alpha\cdot 
(a_{\tilde{\vec{\scriptstyle k}}}+a^*_{-\tilde{\vec{\scriptstyle k}}})
\left[ \cos\tilde{k}_z\tilde{z}-(-1)^m\e^{-\tilde{k}_\perp /2}
\cosh\tilde{k}_\perp\tilde{z} \right]
\right\} \biggr],
\label{eq:deri_a}
\end{eqnarray} 
where $\tilde{k}_{+}=\tilde{k}_x\cos\alpha +\mathrm{i}\tilde{k}_y$ and 
$\tilde{k}_{-}=\tilde{k}_x\cos\alpha -\mathrm{i}\tilde{k}_y$.
For odd modes, $\tilde{k}_z=(2m+1)\pi$ ($m$: integer),
\begin{eqnarray}
 \lefteqn{(\cos\alpha\partial_{\tilde{x}}+\mathrm{i}\partial_{\tilde{y}}
-\sin\alpha\partial_{\tilde{z}})\Phi
=\sum_{\tilde{\vec{\scriptstyle k}}}
\frac{\e^{\mathrm{i}(\tilde{k}_x\tilde{x}+\tilde{k}_y\tilde{y})}}
{\tilde{k}^2}}
\nonumber\\
\times\biggl[  &-& \left\{
-(\tilde{k}_{+}\tilde{k}_{-}a_{\tilde{\vec{\scriptstyle k}}}
-\tilde{k}_{+}^2a^*_{-\tilde{\vec{\scriptstyle k}}})
\left[ \sin\tilde{k}_z\tilde{z}-(-1)^m\e^{-\tilde{k}_\perp /2}
\sinh\tilde{k}_\perp\tilde{z} \right] \right. \nonumber\\
&&-\left. \mathrm{i}\tilde{k}_{+}\tilde{k}_z\sin\alpha\cdot
(a_{\tilde{\vec{\scriptstyle k}}}+a^*_{-\tilde{\vec{\scriptstyle k}}})
\left[ \cos\tilde{k}_z\tilde{z}-(-1)^m\e^{-\tilde{k}_\perp /2}
(\tilde{k}_z /\tilde{k}_\perp )\cosh\tilde{k}_\perp\tilde{z} \right]
\right\} \nonumber\\
&+& \sin\alpha\left\{
\mathrm{i}(\tilde{k}_{-}a_{\tilde{\vec{\scriptstyle k}}}
-\tilde{k}_{+}a^*_{-\tilde{\vec{\scriptstyle k}}})
\left[ \tilde{k}_z\cos\tilde{k}_z\tilde{z}-(-1)^m
\e^{-\tilde{k}_\perp /2}
\tilde{k}_\perp\cosh\tilde{k}_\perp\tilde{z} \right] 
\right. \nonumber\\
&&+\left. \tilde{k}_z^2\sin\alpha\cdot 
(a_{\tilde{\vec{\scriptstyle k}}}-a^*_{-\tilde{\vec{\scriptstyle k}}})
\left[ \sin\tilde{k}_z\tilde{z}-(-1)^m\e^{-\tilde{k}_\perp /2}
\sinh\tilde{k}_\perp\tilde{z} \right]
\right\} \biggr].
\label{eq:deri_b}
\end{eqnarray} 
Let us project Eqs.~(\ref{eq:deri_a}) and (\ref{eq:deri_b}) onto 
$\cos\tilde{k}_z\tilde{z}$ and $\sin\tilde{k}_z\tilde{z}$, respectively.
For $-1/2 < \tilde{z} < 1/2$, 
the projection $F(\tilde{z})$ of a function $f(\tilde{z})$ onto 
$\cos\tilde{k}_z\tilde{z}$ is given by, 
\begin{equation}
 F(\tilde{z})=\int_{-1/2}^{1/2}\d\tilde{z} 
 f(\tilde{z})\cos\tilde{k}_z\tilde{z}
  \left/ \int_{-1/2}^{1/2}\d\tilde{z} \cos^2\tilde{k}_z\tilde{z}.\right.
\end{equation}
After the projection, Eqs.~(\ref{eq:deri_a}) and (\ref{eq:deri_b}) are
reduced to
\begin{eqnarray}
 (\cos\alpha\partial_{\tilde{x}}&+&\mathrm{i}\partial_{\tilde{y}}
-\sin\alpha\partial_{\tilde{z}})\Phi=
\sum_{\tilde{\vec{\scriptstyle k}}}
\frac{\e^{\mathrm{i}(\tilde{k}_x\tilde{x}+\tilde{k}_y\tilde{y})}}{k^2}
\left\{ 1-(2-\delta_{m0})\left( 1-\e^{-\tilde{k}_\perp}\right) 
 \frac{\tilde{k}_\perp}{k^2}\right\} \nonumber\\
&& \times\left\{ (\tilde{k}_{+}\tilde{k}_{-}+\tilde{k}_z^2\sin^2\alpha )
a_{\tilde{\vec{\scriptstyle k}}} 
+ (\tilde{k}_{+}^2+ \tilde{k}_z^2\sin^2\alpha)
a^*_{-\tilde{\vec{\scriptstyle k}}} \right\}\cos\tilde{k}_z\tilde{z},
\end{eqnarray}
where $\delta_{ij}$ is the Kronecker delta, and
\begin{eqnarray}
 (\cos\alpha\partial_{\tilde{x}}&+&\mathrm{i}\partial_{\tilde{y}}
-\sin\alpha\partial_{\tilde{z}})\Phi=
\sum_{\tilde{\vec{\scriptstyle k}}}
\frac{\e^{\mathrm{i}(\tilde{k}_x\tilde{x}+\tilde{k}_y\tilde{y})}}{k^2}
\left\{ 1-2\left( 1-\e^{-\tilde{k}_\perp}\right) 
\frac{\tilde{k}_\perp}{k^2} \right\} \nonumber\\
&& \times\left\{ (\tilde{k}_{+}\tilde{k}_{-}+\tilde{k}_z^2\sin^2\alpha )
a_{\tilde{\vec{\scriptstyle k}}} 
- (\tilde{k}_{+}^2 +\tilde{k}_z^2\sin^2\alpha)
a^*_{-\tilde{\vec{\scriptstyle k}}} \right\}\sin\tilde{k}_z\tilde{z}, 
\end{eqnarray}
respectively. 
Let us impose a restriction on 
the kinds of parameters about wave vectors to describe
the equation of motion: we only use $k$, 
$k_{\perp}=\sqrt{k_x^2+k_y^2}$, $\tilde{k}_z=\pi n$ ($n$: integer), 
$\theta_{\vec{\scriptstyle k}}$ and
$\varphi_{\vec{\scriptstyle k}}$. 
Here, $\sin\theta_{\vec{\scriptstyle k}}=k_{\perp}/k$;
$\cos\varphi_{\vec{\scriptstyle k}}=k_x/k_{\perp}$;
$\sin\varphi_{\vec{\scriptstyle k}}=k_y/k_{\perp}$.
The restriction leads to the following:
\begin{eqnarray}
 \tilde{k}_{+}\tilde{k}_{-}+\tilde{k}_z^2\sin^2\alpha &=&
k_\perp^2 +2 (k_\perp
\cos\varphi_{\tilde{\vec{\scriptstyle k}}} +\tilde{k}_z\sin\alpha)
\tilde{k}_z\sin\alpha;
\nonumber\\
\tilde{k}_{+}^2 +\tilde{k}_z^2\sin^2\alpha &=&
k_\perp^2\e^{2\mathrm{i}\varphi_{\vec{\scriptstyle k}}}
+2 (k_\perp
\e^{\mathrm{i}\varphi_{\vec{\scriptstyle k}}}
+\tilde{k}_z\sin\alpha)\tilde{k}_z\sin\alpha; \nonumber\\
\tilde{k}_\perp &=& \sqrt{k^2-\tilde{k}_z^2}.
\end{eqnarray}
From the above two equations and Eq.~(\ref{eq:LLl2}), we obtain
Eq.~(\ref{eq:al}):
\begin{equation}
\partial_t a_{\vec{\scriptstyle k}}
 +\mathrm{i}(1-\mathrm{i}\lambda) A_{\vec{\scriptstyle k}}
   a_{\vec{\scriptstyle k}}
 +\mathrm{i}(1-\mathrm{i}\lambda) B_{\vec{\scriptstyle k}}
   a^*_{-\vec{\scriptstyle k}}=0. 
\end{equation} 
For even modes,
\begin{equation}
 A_{\vec{\scriptstyle k}}= C_{\vec{\scriptstyle k}}+ 
 \mathrm{Re}\tilde{C_{\vec{\scriptstyle k}}}, \quad
 B_{\vec{\scriptstyle k}}= \left(
D_{\vec{\scriptstyle k}}
\e^{2\mathrm{i}\varphi_{\vec{\scriptscriptstyle k}}}
+\tilde{C_{\vec{\scriptstyle k}}}
\right),  
\label{eq:ABa}
\end{equation}
and for odd modes,
\begin{equation}
 A_{\vec{\scriptstyle k}}= C_{\vec{\scriptstyle k}}+ 
 \mathrm{Re}\tilde{C_{\vec{\scriptstyle k}}}, \quad
 B_{\vec{\scriptstyle k}}= -\left(
D_{\vec{\scriptstyle k}}
\e^{2\mathrm{i}\varphi_{\vec{\scriptscriptstyle k}}}
+\tilde{C_{\vec{\scriptstyle k}}}
\right).  
\label{eq:ABb}
\end{equation}
Here we note that 
 $C_{\vec{\scriptstyle k}}$, $\tilde{C_{\vec{\scriptstyle k}}}$
and $D_{\vec{\scriptstyle k}}$ are described in Eq.~(\ref{eq:CDf}).
The difference between Eqs.~(\ref{eq:ABa}) and (\ref{eq:ABb}) is just
the sign of $B_{\vec{\scriptstyle k}}$. Therefore, for both even and odd
 modes, we may write
\begin{equation}
 B_{\vec{\scriptstyle k}}= (-1)^n\left(
D_{\vec{\scriptstyle k}}
\e^{2\mathrm{i}\varphi_{\vec{\scriptscriptstyle k}}}
+\tilde{C_{\vec{\scriptstyle k}}}
\right). 
\end{equation}


\begin{thebibliography}{99}
 \bibitem{Schlomann} E.~Schl\"omann, J.~J.~Green and U.~Milano,
		     J. Appl. Phys. {\bf 31}, 386S (1960).
 \bibitem{Suhl} H.~Suhl, J. Phys. Chem. Solids {\bf 1}, 209 (1957).
 \bibitem{Patton} See, for example, M.~Chen and C.~E.~Patton, 
 in {\it Nonlinear Phenomena and Chaos in Magnetic Materials}, edited by
	 P.~E.~Wigen (World Scientific, Singapore, 1994), pp. 33-82.
 \bibitem{Kalin} B.~A.~Kalinikos, N.~G.~Kovshikov and N.~V.~Kozhus,
	 Sov. Phys. Solid State {\bf 27}, 1681 (1986). [Fiz. Tverd. Tela
	 (Leningrad) {\bf 27}, 2794 (1985).]
 \bibitem{Wiese} G.~Wiese, L.~Buxmzn, P.~Kabos and C.~E.~Patton,
	 J.Appl. Phys. {\bf 75}, 1041 (1994).
 \bibitem{Kabos} P.~Kabos, M~Mendik, G.~Wiese and C.~E.~Patton,
	 Phys. Rev. B {\bf 55}, 11457 (1997).
 \bibitem{Laksh} M.~Lakshmanan and K.~Nakamura, Phys. Rev. Lett. 
	 {\bf 53}, 2497 (1984). 
 \bibitem{Elmer} F.~J.~Elmer, Phys. Rev. B {\bf 53}, 14323 (1996).
 \bibitem{Kudo} K.~Kudo and K.~Nakamura, in preparation.
\end{thebibliography}
\end{document}